\documentclass[sigconf,language=english]{acmart}
\AtBeginDocument{%
  \providecommand\BibTeX{{%
    \normalfont B\kern-0.5em{\scshape i\kern-0.25em b}\kern-0.8em\TeX}}}
\usepackage{amsmath,amsfonts}
\usepackage{algorithm}
\usepackage{algorithmicx}
\usepackage{algpseudocode}

\usepackage{graphicx}
\usepackage{svg}
\usepackage{subcaption}
\usepackage{tikz}
\usepackage[export]{adjustbox}
\usepackage{pdfpages}
\usepackage{textcomp}
\usepackage{xcolor}
\def\BibTeX{{\rm B\kern-.05em{\sc i\kern-.025em b}\kern-.08em
    T\kern-.1667em\lower.7ex\hbox{E}\kern-.125emX}}

\newcommand{\todo}[1]{{\color{black}{#1}}}
\usepackage{ulem}
\usepackage{qcircuit}

\usepackage{enumitem}
\setlist{noitemsep,left=0pt,topsep=0pt,parsep=0pt,partopsep=0pt}
\usepackage[small,compact,noindentafter]{titlesec}
\titlespacing\section{0pt}{6pt plus 2pt minus 2pt}{2pt plus 2pt minus 2pt}
\titlespacing\subsection{0pt}{4pt plus 2pt minus 2pt}{2pt plus 1pt minus 1pt}
\titlespacing\subsubsection{0pt}{2pt plus 0pt minus 1pt}{1.0ex}
\titlespacing{\paragraph}{0pt}{2pt plus 0pt minus 1pt}{1.0ex}

\copyrightyear{2024}
\acmYear{2024}
\setcopyright{rightsretained}
\acmConference[ICCAD '24]{IEEE/ACM International Conference on
Computer-Aided Design}{October 27--31, 2024}{New York, NY, USA}
\acmBooktitle{IEEE/ACM International Conference on Computer-Aided Design
(ICCAD '24), October 27--31, 2024, New York, NY, USA}
\acmDOI{10.1145/3676536.3676719}
\acmISBN{979-8-4007-1077-3/24/10}

\begin{document}

%\title{A Hardware-Aware Framework \break for Practical Quantum Circuit Knitting}
\title[A Hardware-Aware Gate Cutting Framework for Practical Quantum Circuit Knitting]{A Hardware-Aware Gate Cutting Framework for Practical Quantum Circuit Knitting}

\author{Xiangyu Ren}
\authornote{Xiangyu Ren has started this work during an internship at Tencent Quantum Lab.}
\affiliation{
  \institution{The University of Edinburgh}
  \city{Edinburgh}
  \country{United Kingdom}
}
\email{xiangyu.ren@ed.ac.uk}

\author{Mengyu Zhang}
\affiliation{
  \institution{Tencent Quantum Lab}
  \city{Shenzhen}
  \country{China}
}
\email{mengyuzhang@tencent.com}

\author{Antonio Barbalace}
\affiliation{
  \institution{The University of Edinburgh}
  \city{Edinburgh}
  \country{United Kingdom}
}
\email{abarbala@ed.ac.uk}

\begin{abstract}
Circuit knitting emerges as a promising technique to overcome the limitation of the few physical qubits in near-term quantum hardware by cutting large quantum circuits into smaller subcircuits. Recent research in this area has been primarily oriented towards reducing subcircuit sampling overhead. Unfortunately, these works neglect hardware information during circuit cutting, thus posing significant challenges to the follow on stages. In fact, direct compilation and execution of these partitioned subcircuits yields low-fidelity results, highlighting the need for a more holistic optimization strategy.

In this work, we propose a hardware-aware framework aiming to advance the practicability of circuit knitting. Drawing a contrast with prior methodologies, the presented framework designs a cutting scheme that concurrently optimizes the number of gate cuttings and SWAP insertions during circuit cutting. In particular, we leverage the graph similarity between qubits interactions and chip layout as a heuristic guide to reduces potential SWAPs in the subsequent step of qubit routing. Building upon this, the circuit knitting framework we developed has been evaluated on several quantum algorithms, leading to reduction of total subcircuits depth by up to 64\% (48\% on average) compared to the state-of-the-art approach, and enhancing the relative fidelity up to 2.7$\times$. 
\end{abstract}

% \keywords{Quantum computing, Circuit knitting, Hardware-Aware, Circuit cutting, Routing overhead}

\maketitle

\section{Introduction}

Quantum computing has %holds
the potential to solve many classically intractable problems. However, the limited number of available qubits in current Noisy Intermediate Scale Quantum (NISQ) devices hinders quantum computing from realizing its initially anticipated computational power.

\begin{figure}[!t]
% \vspace{3mm}
    \centering
    \includegraphics[page=1,width=.47\textwidth]{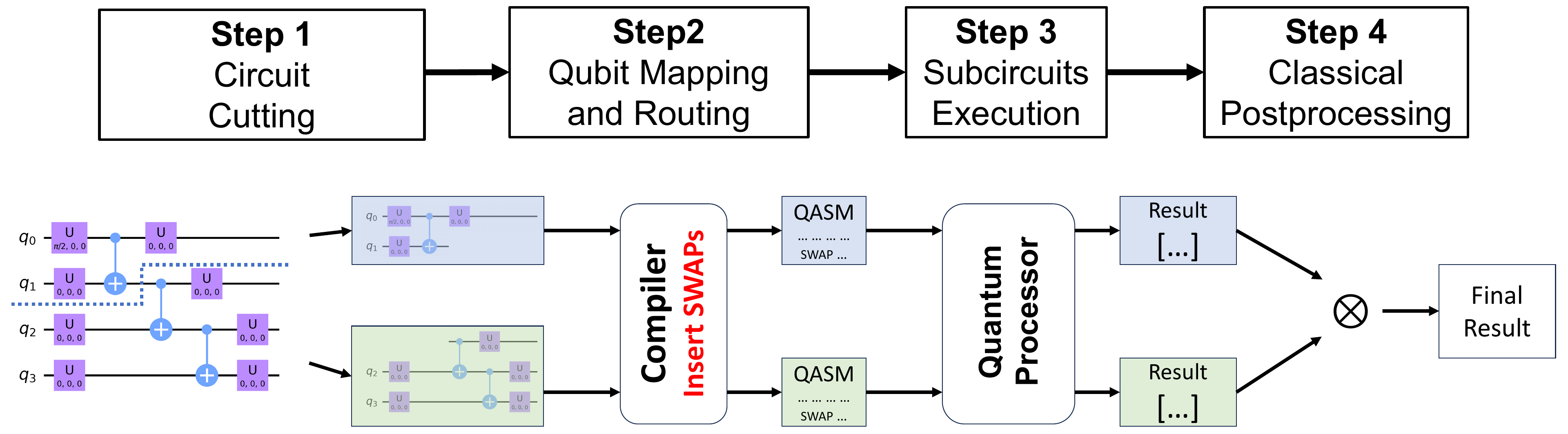}
    \caption{Generalization of previous knitting frameworks.} 
    \label{fig:oldframework}
\vspace{-3mm}
\end{figure}

Quantum circuit knitting~\cite{bravyi2016, peng2020, mitarai2021, Piveteau2023knitting} has been put forward as a technique to solve this problem by running large circuits with more qubits than physically available. Fig.~\ref{fig:oldframework} shows a generalization of previous circuit knitting frameworks, e.g., CutQC~\cite{cutqc}. We have our input circuit as OpenQASM format. First, the original circuit is cut along qubit wires (wire cutting) or 2-qubit gates (gate cutting, not in Figure) to form separated subcircuits. Then, subcircuits are individually compiled and executed on quantum processors. Finally, %In the final step,
classical postprocessing: individual execution results are reconstructed into the result of the original circuit.

%However, the cost of this technique is a subcircuits sampling overhead that scales exponentially in the number of cuts \cite{Piveteau2023knitting}.
However, this technique is costly as the overhead of subcircuits sampling (step 3) and classical postprocessing (step 4) scale exponentially with the number of cuts~\cite{Piveteau2023knitting}.
Hence, to lower such overheads, previous research focuses on minimizing the number of cuts~\cite{cutqc,tomesh2021divide}. 
Other research involving cluster-based hardware~\cite{baker2020} also focus on minimizing the number of cuts.
% \todo{cluster-based hardware?}

\begin{figure}[b]
    \centering
    \includesvg[width=0.47\textwidth]{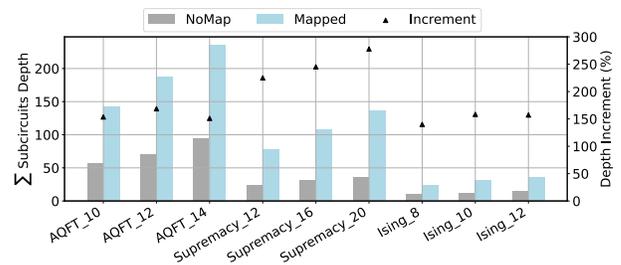}
    \caption{Example of qubit routing overhead, measured by the sum of depth of each subcircuits. "NoMap" shows compilation results with all-to-all qubit connectivity (or before compilation), "Mapped" shows the results when subcircuits are compiled with IBM Lagos (7-qubit) as backend. "Increment" shows the percentage of increased $\sum Subcircuit ~ Depth$ from "NoMap" to "Mapped", with its axis on the right.}
    % ORGINAL Example of qubit routing overhead. "NoMap" represents the compilation results when assuming all-to-all qubit connectivity, "Mapped" shows the results when subcircuits are compiled with IBM Lagos (7-qubit) as backend.}
    \label{fig:introexample}
\vspace{-3mm}
\end{figure}

\begin{figure*}[t]
    \centering
    \includegraphics[page=1,width=.97\textwidth]{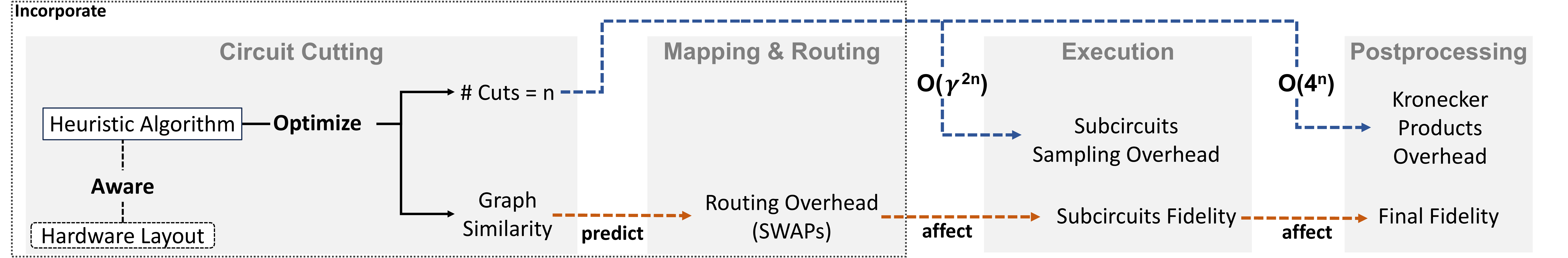}
    \caption{Optimization structure inside our framework. The \textit{number of cuts} affects the sample overhead and postprocessing overhead with an exponential relationship (blue arrows), while \textit{graph similarity} predicts the routing overhead (orange arrows). Being aware of the hardware layout, our algorithm optimize both of them.}
    \label{fig:framework}
% \vspace{-3mm}
\end{figure*}

Despite their efficiency in minimizing the number of cuts (step 1), %sampling and postprocessing overhead during step 1, %%AB: I understand what you mean but it doesn't sound right in English -- anyway, feel free to do what you want
the follow-on compilation (step 2) may counteract the improvements. 
This is because during compilation a %n excessive 
large number of SWAPs may be inserted in the qubit routing stage, which likely extends the depth of each subcircuit and undermines the fidelity. \todo{In Qiskit, this is transpilation. }
Fig.~\ref{fig:introexample} shows a comparison of the subcircuits depth before and after compilation. %qubit mapping \& routing. 
%The cause of this problem is that neglecting hardware information during circuit cutting make partitioned subcircuits challenging to route onto actual devices.
%In previous frameworks, the SWAPs insertion remains unknown till we finish the entire circuit cutting procedure and start compilation, which made the final result vulnerable to mapping and routing.
The depth increase is caused by neglecting hardware information during circuit cutting, which makes the routing of the resulting subcircuits onto actual devices challenging.
In fact, in previous works, the SWAPs insertion remains unknown till we finish the circuit cutting procedure and start compilation, which makes the final result vulnerable to mapping and routing overheads. 

%In step 2, the aforementioned research directly compile the subcircuits towards specific hardware. 
% It can lead to severe consequences, as neglecting hardware information during circuit cutting make partitioned subcircuits challenging to route onto actual devices, subsequently introducing an excessive number of inserted SWAPs in the final result.
%During qubit routing stage in compilation, inserting SWAPs extend the depth of each subcircuit and significantly undermine the fidelity. 

% It is obvious to see that finding a proper circuit cutting is vital to improving the circuit depth and fidelity in circuit knitting tasks.

\paragraph{\textbf{Hardware-aware Circuit Cutting.}}
Instead of investigating new qubit mapping and routing methods %, which have been already
-- widely studied in recent years, this work proposes to \textit{incorporate} the routing problem (step 2) into the circuit cutting (step 1). 
This is in sharp contrast to previous works, where such steps execute sequentially, and independently.
Our idea is to improve the quality of circuit cutting by considering the routing overhead on the actual hardware during cutting. 

Based on this idea, we exploit the fact that graph similarity between \textbf{a subcircuit's qubit interaction graph} and \textbf{hardware layout} can well predict the number of SWAPs for qubit routing. Setting graph similarity as an optimization objective, we can iteratively improve the quality of circuit cutting. 
Our hardware-aware circuit knitting framework is depicted in Fig.~\ref{fig:framework}, including relationships and costs for each step. These steps are briefly detailed below.

\begin{enumerate}[label=\alph*)]
%\noindent
 \item \textbf{Circuit Cutting:} %Commencing with circuit cutting as the first step, 
 Our algorithm co-optimizes two objectives: 1) minimizing the number of gate cuts, 2) maximizing the subcircuit graph similarity to the hardware-layout of a quantum processor.
% From the framework level,
Graph similarity serves as a "heuristic clue" to reduce SWAPs before the qubit mapping \& routing step, as shown in the left half of Fig.~\ref{fig:framework}. After dividing logical qubits into different subcircuits, we insert those local single-qubit operations for replacement using Qiskit.

%\noindent
 \item \textbf{Compilation and Execution:} Each subcircuit is compiled through qubit mapping \& routing, where SWAPs insertion are actually performed. Then subcircuits are individually executed on one or more quantum processor. This step is unchanged in our framework, easily integrating in existent frameworks.

%\noindent
 \item \textbf{Classical Postprocessing:} Finally, we also implement an efficient module for reconstructing results during classical postprocessing. The classical postprocessing can be represented by \textit{tensor network contraction}, and a contraction tree optimizer is utilized to find the optimal sequence for contraction.
% We identify an optimization opportunity to reduce the Kronecker products under gate cut protocol by utilizing an optimized tree-based order to perform products computation
\end{enumerate}

Apart from quantum circuit knitting, our design %insight
is also beneficial for cutting circuits in other distributed quantum computing scenarios~\cite{autocomm}, e.g. chiplet architecture~\cite{chiplet}.
\paragraph{\textbf{Contributions.}}
Our key contributions can be summarized as: % follows:

\begin{enumerate}
  \item We \emph{introduce} a new quantum circuit knitting framework design where the circuit cut stage exploits the quantum hardware-layout to make partitioning decisions, %this is different from cut and compiler co-design
  while maintaining compatibility with existent frameworks. \todo{This work is the first to consider and co-optimize three distinct types of overhead in circuit knitting tasks, as detailed in Section 2.2;}
  \item We specialize such design for the \textit{gate cut} technique -- already demonstrated in theory, and \emph{introduce} a new graph-similarity-based approach to minimize the number of gates to be cut while reducing SWAPs in subcircuits by exploiting the hardware-layout;
  % \item We further \emph{introduce} a new classical postprocessing algorithm, targeting the gate cut technique, inspired by tensor-contraction-tree;
  \item We fully \emph{implemented} the entire framework based on Qiskit, \emph{deployed} it on a server with GPU, and \emph{run} subcircuits on IBM quantum processors. We evaluated and compared our framework with the state-of-the-art work CutQC. When running benchmark where the original qubit count is $1.5\times$ -- $4\times$ of physical qubits on the chip, our work reduce the subcircuits depth by 48\% compared to CutQC. And it shows the relative fidelity up to $2.7\times$ ($1.67\times$ on average). 
%%ORIGINAL:
%    \item We disclose a graph-similarity-based circuit cutting optimization, then design an algorithm that can significantly reduce the subcircuits depth after compilation, thus improve the fidelity in a circuit knitting task by $3\times$ on average.

    % \item We also propose a postprocessing approach inspired by tensor-contraction-tree, which can accelerate postprocessing and tackling the exponential overhead of classical computation by $\times$ times.

%    \item Our research takes into consideration three critical factors that affect efficiency and quality of a circuit knitting task: 1) subcircuit fidelity, 2) subcircuit sampling overhead and 3) classical postprocessing overhead, thus providing a practical framework for circuit knitting.
\end{enumerate}

%=========================================================================================
\section{Problem Analysis}
% In this section, we start with an overview of fundamental works on circuit knitting. Then, we analyse the key factor that impacting the overheads in circuit knitting, emphasizing the importance of our optimization objectives: \#Cuts and \#SWAPs.
\subsection{Background} 
Quantum circuit knitting aims to address the constraint of limited physical qubits on NISQ-era quantum computers. 
The theory behind circuit knitting is quasiprobability simulation \cite{pashayan2015}: the expected value of the original circuit is obtained by sampling each smaller part of it that we define as subcircuit. Based on the quasiprobability simulation, previous works provide several approaches to partition the quantum circuit into smaller ones. Specifically, this is realized by decomposing a \textit{qubit wire} or a \textit{2-qubit gate} into a set of \textit{single qubit operations}, followed by classical postprocessing to reconstruct the expectation value of the original circuit. 

\begin{figure}[ht]
\centering
\resizebox{1\linewidth}{!}{
    \Qcircuit @C=1em @R=1em {
    & \ctrl{1}  & \gate{R_Z(\frac{\pi}{2})} & \qwa  & & & \qw   & \qwa  &&& \gate{Z} &  \qwa & & & & & & & & & & &  \meterB{{}_{=\alpha_1}}                      & \qwa & & & \gate{R_Z(\frac{\alpha_1\pi}{2})}  & \qwa       \\ 
    &           &                           &       & = &       &       & & +       & & & & & & & & + \displaystyle\sum_{\alpha_1 , \alpha_2 \in \{\pm 1\}^2}^{} \alpha_1 \alpha_2 \Bigg\{ & & & & & & & & + & & & & \Bigg\}  \\
    & \ctrl{-1} & \gate{R_Z(\frac{\pi}{2})} & \qwa  & & & \qw   & \qwa  &&& \gate{Z} &  \qwa & & & & & & & & & & &  \gate{R_Z(\frac{\alpha_2\pi}{2})} & \qwa & & & \meterB{{}_{=\alpha_2}}                       & \qwa       }
}
    \caption{Example of CZ gate cut decomposition. There are four terms of single qubit operations, and each term represents a substitution for the original CZ gate. While the last two terms are associate with $\alpha_i$ coefficients, the mid-circuit measurement boxes stand for "post selection", i.e., the result will be kept only if "mid-circuit measurement result == $\alpha_i$".}
    \label{fig:v2qg}
    \vspace{-5mm}
\end{figure}
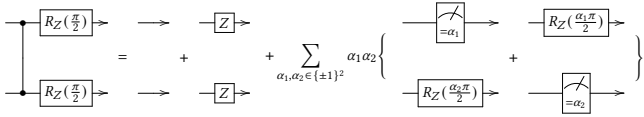

\paragraph{\textbf{Theoretical Works.}} 
Peng et al. proposed \textit{wire cut}~\cite{peng2020} and proved its validity. Later, Mitara et al. introduced \textit{gate cut}~\cite{mitarai2021} that is theoretically established to have significantly lower sampling overhead than \textit{wire cut}~\cite{Piveteau2023knitting}. Along with \textit{entanglement forging}~\cite{Eddins2022forging}, these techniques are known as \textit{circuit knitting}~\cite{Piveteau2023knitting}. Fig.~\ref{fig:v2qg} shows the CZ gate being decomposed into an expectation combination of multiple terms of single qubit operations, while in each term the control and target qubits are substituted by single qubit gates and mid-circuit measurements. 

\paragraph{\textbf{Practical Implementations.}}
Tang et al. proposed the CutQC \textit{wire cut} framework~\cite{cutqc}, with minimizing the classical postprocessing overhead as optimization target, followed by their work ScaleQC further improving the postprocessing of \textit{wire cut}~\cite{scaleqc}. 
Brandhofer et al. incorporate \textit{wire cut} and \textit{gate cut} to minimize the sampling overhead~\cite{hybridcut}, while Pawar et al. additionally consider qubit reuse in their work~\cite{pawar2023integrated}.
Chen et al. explored a fast approach for classical postprocessing of \textit{wire cut}~\cite{approx}, and Tomesh et al. apply the \textit{wire cut} to solve Maximum Independent Set (MIS) problem~\cite{tomesh2021divide}. 
\todo{
%ORIGINAL However, it is important to note that these studies did not take qubit routing overhead into consideration.
While none have addressed the problem of incorporating \textit{gate cut} into a practical framework before, 
it is important to note that these studies did not take qubit routing overhead into consideration.
}

\paragraph{\textbf{Wire Cut and Gate Cut.}}
\todo{We have developed our framework using the \textit{gate cut} technique rather than the \textit{wire cut} technique or a hybrid of both, because the \textit{gate cut} protocol is more concisely represented in the graph model (detailed in Section 3.1). This makes it more intuitive for verifying our concept. Additionally, the \textit{gate cut} approach has the potential to benefit from enhancements through classical communication~\cite{Piveteau2023knitting}, which holds promise for future distributed quantum computing.}
Note that the gate decomposition protocol in our framework can be switched to a circuit knitting scheme based on classical communication with minor modifications. However, this approach warrants further exploration in conjunction with more mature and advanced quantum control systems in the near future.

\subsection{Practical Circuit Knitting \emph{Overheads}}\label{sec:overheads}

%ORIGINAL Existent approaches to circuit knitting are affected by at least the followings overheads that make them non-practical in many cases. Further, they depends on the number of SWAPs and cuttings.
Existent approaches to circuit knitting are in many cases non-practical due to at least the following overheads, which depends on the number of SWAPs and the number of cuttings.

\paragraph{\textbf{Circuit Depth and Routing Overhead (depends on \emph{\#SWAPs}}).} %In the example we provided in Fig~\ref{fig:introexample}, we quantify
In Fig.~\ref{fig:introexample}, we report the sum of subcircuits depth before and after the qubit mapping and routing (step 2, in Fig.~\ref{fig:oldframework}) targeting a specific hardware.
From the graph, we can see that qubit mapping and routing increase the circuit depth from $1.4\times$ to $2.8\times$, this is due to the SWAP insertions, each including 3 CNOT gates, which account for a large portion of each subcircuit after compilation.
A deeper circuit will significantly lower the fidelity of subcircuit, and finally affect the fidelity of original circuit, we represent this relation with the bottom brown arrow in Fig.~\ref{fig:framework}.
%Corresponding to the bottom arrows in  the expanded circuit depth would significantly lower the fidelity of subcircuits, 
Therefore, minimizing the routing overhead is critical to improve the fidelity, and the routing overhead is hardware dependent.

\paragraph{\textbf{Subcircuits Sampling Overhead (depends on \emph{\#cuttings}).}} Sampling overhead is the number of shots each subcircuit needs to run in order to grantee the accuracy of circuit knitting. Generally, the sampling overhead $O(\gamma^{2n})$ scales exponential with the number of cuts $n$, while $\gamma=3$ for \textit{gate cut} and $\gamma=4$ for \textit{wire cut} \cite{Piveteau2023knitting,wirecutcomm}. 
Thus, the lower $\gamma$ reinforces our decision of selecting \textit{gate cut} protocol, and minimizing the number of cuts $n$ is important to keep the sampling overhead low.

\paragraph{\textbf{Classical Postprocessing Overhead (depends on \emph{\#cuttings}).}} 
In the classical postprocessing step, the result of the original circuit is reconstructed through Kronecker products. %Prior research
State-of-the-art estimates $O(4^n)$ Kronecker products for computation \textit{wire cut} framework, where $n$ is the number of cuts~\cite{cutqc,tomesh2021divide}.
With our implementation based on tensor network contraction, the Kronecker products computation is also $O(4^n)$ (detailed in Section~4), the same as \textit{wire cut}. Hence, we will focus on comparing the number of cuts $n$. %compare the postprocessing overhead with previous work by directly comparing number of cuts $n$.

\subsection{Motivation: A Case Study}

\begin{figure}[t]
    \centering
    \includegraphics[page=1,width=.4\textwidth]{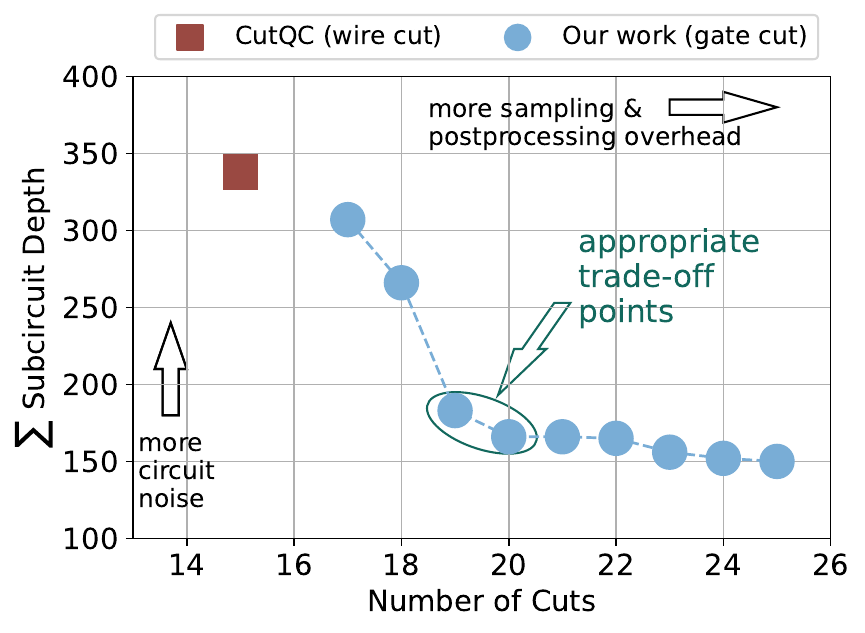}
    \caption{Balancing between different overheads in circuit knitting. In this motivation experiment, we cut and compile a AQFT\_16 circuit to the IBM Lagos (7-qubit) quantum hardware. Y-axis shows the sum of subcircuit depth, representing the routing overhead; while X-axis shows the number of cuts related to sampling \& postprocessing overhead.} 
    \label{fig:motivation}
    \vspace{-2mm}
\end{figure}

\todo{Based on the various types of overheads discussed in Section 2.2, our objective is to explore the correlations and trade-offs between them in practical scenarios. Fig.~\ref{fig:motivation} illustrates our analysis by showcasing a search for a circuit cutting solution on one of our benchmark circuits. While the CutQC method provides an optimal solution that minimizes the number of wire cuts (marked by red square scatter points), our framework operates along the Pareto frontier, balancing between qubit routing overhead and sampling/postprocessing overhead (represented by blue circle scatter points). It is important to note that while our solution may not precisely align with the optimal Pareto frontier, it effectively demonstrates the general trends and trade-offs between these overheads. We observe that an appropriate trade-off can lead to significantly improved circuit fidelity.}

\todo{However, identifying the optimal trade-offs during the circuit cutting of arbitrary quantum circuits is not straightforward. Consequently, there is a pressing need for a strategy that can autonomously determine suitable solutions. In Section 3, we will introduce an algorithm designed to systematically search for circuit cutting solutions that achieve a balanced trade-off between the involved overheads.}

%=============================================================================
\section{Framework Design}
\subsection{Modelling the Circuit Cutting Problem}
\begin{figure*}[t]
    \includegraphics[page=1,width=0.85\textwidth,center]{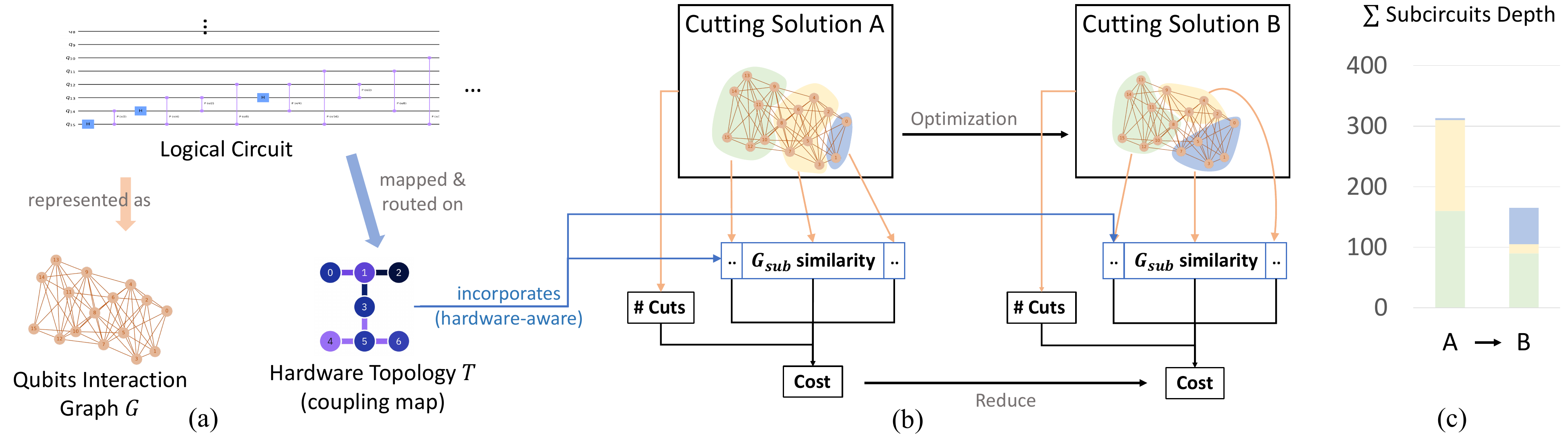}
    \caption{Hardware-aware optimization for circuit cutting. (a) An example of the qubit interaction graph $G$, with AQFT\_16 as the original logical circuit. (Every $W_{e_{ij}}$ has value 1 in this circuit, so we don't draw it on $G$). The circuit will be cut and compiled to IBM Lagos (hardware topology $T$). (b) Optimizing the cutting solution with multiple objectives: \#Cuts and Graph Similarity, lead to a reduction in subcircuits depth, shown in (c).}
    \label{fig:layoutexample}
\end{figure*}

%To reason about various circuit cuttings, first we need a more abstracted model to represent the process that original circuit being cut into subcircuits.
To reason about the best partitioning of a quantum circuit into subcircuits we decided to use a more abstract model than the circuit itself: we use the qubit interaction graph~\cite{characterization}. % to represent the circuit.
In the qubit interaction graph $G=(V,E)$, each logical qubit $l_i$ is represented by a vertex $v_i \in V$, while 2-qubit gate $g_{ij}$ performed between logical qubits $l_i$ and $l_j$ is represented as edge $e_{ij} \in E$. The weight $W_{e_{ij}}$ of each edge indicates the number of 2-qubit gates applied between the corresponding pair of logical qubits $l_i$ and $l_j$. An illustrative example of the graph is presented in Fig.~\ref{fig:layoutexample}(a). If the circuit includes multi-qubit gates, these are first transformed into combinations of single- and two-qubit gates prior to graph generation.

Adopting the qubit interaction graph $G=(V,E)$, gate cutting of a circuit corresponds to partitioning $G$ into subgraphs $G_{sub}$ (upper part of Fig.~\ref{fig:layoutexample}(b)). 
Under these circumstances, the number of cuts is determined by the sum of the weights of the edges that connect between different subgraphs, denoted as:
\begin{equation}\label{eq:cuts}
    \#Cuts = \sum_{v_{i,j}\in V}{W_{e_{ij}} \left[ G_{sub}(v_i) \neq G_{sub}(v_j)\right] }
\end{equation}
Where $G_{sub}(v_i)$ specifically refers to the subgraph that contains $v_i$.
On the other hand, routing overhead is associated solely with the two-qubit gates located within a single subgraph, which are referred to as intra-subgraph edges.

\subsection{Overview of Optimization Process}
Fig.~\ref{fig:layoutexample} shows the general idea of our optimization. In upper part of Fig.~\ref{fig:layoutexample}(b) are two different cutting solutions A and B of the same qubit interaction graph, derived from the same original logical circuit (upper part of Fig.~\ref{fig:layoutexample}).
We can see in Fig.~\ref{fig:layoutexample}(c) that sum of subcircuits depth is largely different between this two solutions after qubit routing.
So, an approach to finding a circuit cut with the minimal number of cuts and SWAPs is sought. 

For \#cuts, it is easy to calculate with Equation.~\ref{eq:cuts}. 
However, for \#SWAPs (or routing overhead), it remains unknown till finishing qubit mapping \& routing.
This requires a circuit cutting algorithm aware of the hardware, hence, the routing overhead. 
\todo{We leverage graph similarity to estimate the routing overhead, as illustrated by the blue arrows connecting between Fig.~\ref{fig:layoutexample} (a) and (b). Details are explained in the following subsections.}

%\subsection{Predicting the \# of SWAPs}
\subsection{Exploiting Hardware Layout}
Finding a cut with minimal gates cut \textbf{and} minimal SWAPs is non-trivial.
This is because we cannot normally infer the number of SWAPs during circuit cutting, but the subcircuits need to be compiled first.
Indeed, we could compile all subcircuits for each possible cutting solution, but this is unrealistic due to the prolonged time for gate cut decomposition and compilation.
Also, the number of SWAPs cannot be defined as a linear function on the circuit; thus, impossible to integrate in a mathematical optimization solver -- alike~\cite{cutqc}.

%\subsection{Use Graph Similarity to Predict \# SWAPs}
\paragraph{\textbf{Graph-Similarity to Predict SWAP insertions.}}
It is vital to explore an approach to estimate SWAP insertions in advance, and measuring graph similarity between \textbf{a subcircuit's interaction graph $G_{sub}$} and \textbf{hardware layout $T$} makes an ideal option. 
Graph similarity quantifies how closely two graphs resemble each other~\cite{graphsim}. 
There are several approaches to quantify the graph similarity, and we identify Graph Edit Distance (GED)~\cite{editdistance} as the most appropriate for our scenario, because of its clarity to describe the difficulty for qubit routing on NISQ scale circuits.

Specifically, we calculate the edit distance from the hardware layout to a subcircuit's interaction graph, denoted as $C_{GED}(G_{sub},T)$, and use this metric to estimate the SWAPs insertion.
SWAP are inserted because of the limited connectivity of physical qubits in the  hardware layout. If we have to add more edges to edit hardware layout into interaction graph (higher edit distance), that means the lack of connectivity is more severe, thus more SWAP have to be added to fill the gap. % (more SWAP insertions).
In the graph edit distance calculation function we only set the cost for adding edges to hardware layout graph, and there's no cost for deleting edges or nodes from hardware layout. 

\begin{figure}[b]
    \begin{subfigure}{0.96\columnwidth}
        \centering
        \includegraphics[width=1\columnwidth]{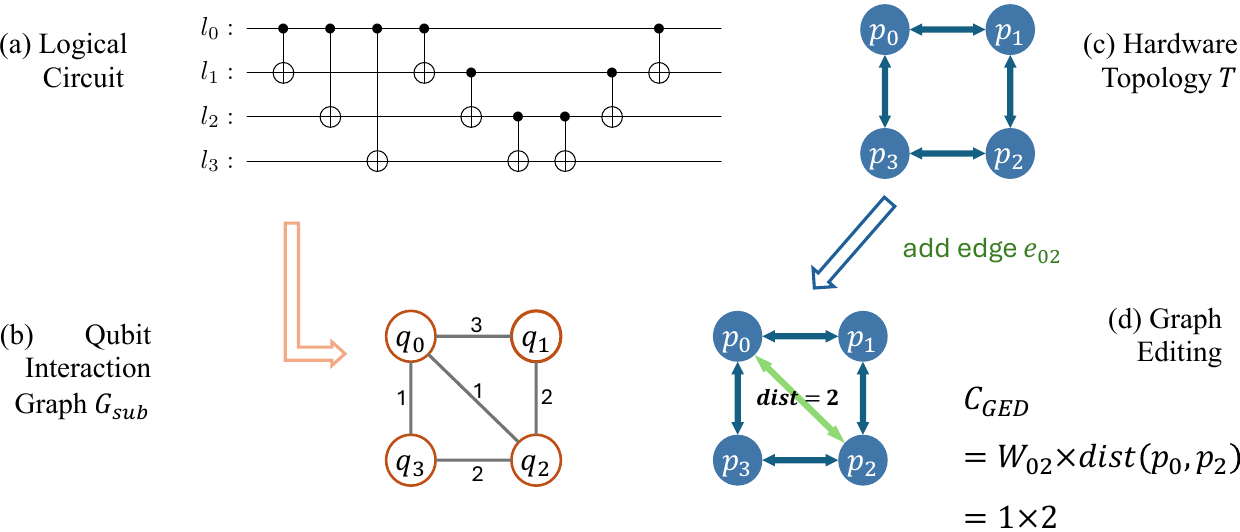}
    \end{subfigure}
    % \vspace{4mm}
    \begin{subfigure}{0.95\columnwidth}
        \centering
        \includesvg[width=1\columnwidth]{fig_correlation.svg}
    \end{subfigure}

    \caption{(a) An example of calculating the graph editing cost. (b) Distribution of the number of SWAPs (Y-axis) for different heuristic cost $C$ (X-axis). We randomly sample 1000 subcircuits cut down from the AQFT\_16 benchmark, and compile them to IBM Lagos. We can see a linear correlation in this case between edit distance and SWAPs.}
    \label{fig:cost}
\end{figure}

\subsection{Designing the Searching Heuristic}
\todo{Building on the concept of Graph Edit Distance introduced in Section~3.3, we propose a heuristic cost function designed to optimize the search for circuit cutting solutions. First, we introduce the \textit{topology distance} (denoted as $dist(p_i,p_j)$), which measures the shortest path between two physical qubits $p_i$ and $p_j$. Each edge in hardware topology is assigned a distance of 1, and $dist(p_i,p_j)$ depict the number of edges in the shortest path connecting these qubits (see Ref.~\cite{sabre} Section~4.1).}

Then, we calculate the graph edit distance utilizing edge weight $W_{ij}$ in interaction graph $G_{sub}$ and topology distance $dist(pi,pj)$ in hardware topology $T$. Let $e_{ij}$ be a edge that should be add to $T$, than the cost for this graph editing is
\begin{equation}\label{eq:ged}
    C_{GED}(G_{sub},T) = \sum_{e_{ij} \in E_{add}}{W_{ij} \times dist(p_i,p_j)}
\end{equation}
%, with 
Where $E_{add}$ be the edge to be added. An example is given in Fig.~\ref{fig:cost}~(a) to help understand the calculation of cost.

To evaluate the effectiveness of our heuristic, we compute the correlation between heuristic cost $C_{GED}$ and SWAP counts on a broad range of circuits, and report AQFT\_16 benchmark in Fig~\ref{fig:cost}~(b). Results confirm our hypothesis.
% \footnote{All graphs will be released as a technical report upon paper acceptance.}.

Finally, we combine \#Cuts and $\sum C_{GED}$ of all subcircuits to construct the overhead heuristic for the circuit cutting solution:
\begin{equation}\label{eq:cost}
    C = \#Cuts + \alpha \cdot \sum C_{GED}
\end{equation}
%, while
Where $\alpha$ is a parameter to balance between these two co-optimizing objectives. \todo{The parameter $\alpha$ is set uniquely for each circuit according to its characteristics (e.g. gate density~\cite{queko}), and it can be adjusted to satisfy varying degrees of trade-off between overheads. A systematic approach for determining the value of $\alpha$ is given in Section~5.1.}

\algdef{SE}[REPEATN]{RepeatN}{End}[1]{\algorithmicrepeat\ #1 \textbf{times}}{\algorithmicend}

\begin{algorithm}[b]
	\caption{Circuit Cutting Algorithm}
	\label{rand}
	\begin{algorithmic}[1]
		\Require Original qubit interaction graph $G_0$, hardware layout of backend $T$, objective weight coefficient $\alpha$, search iterations $r$.
		\Ensure Partitioning solution $\{G_{sol}\}$.
            \State $C_{min} \gets \infty$
		\RepeatN{$r$}
                \State $(C,\{G_{sub}\}) \gets \Call{Merge\_Nodes}{G_0}$
                \If{$C<C_{min}$}
                    \State $C_{min} \gets C$
                    \State $\{G_{sol}\} \gets \{G_{sub}\}$
                \EndIf
            \End
            %function
		\Function{Merge\_Nodes}{interaction graph $G=(V,E)$}
                \State $E_c \gets $ edges connecting between subgraphs
                \State $E_{cst} \gets$ edges in $E_c$ that comply constrain
                \If {$E_{cst} = \varnothing$}
                    \State $C = \#Cuts + \alpha \cdot \sum C_{GED}(G_{sub},T)$
                    \State \Return $(\{G_{sub}\},C)$
                \EndIf
                \State $n \gets |V|$
                \While {$|V| > \sqrt{n}$}
                    \State $e \gets$ random sample from $E_{cst}$
                    % \State Do contraction on a random edge $e \in E_{constr}$
                    \State $G' \gets G$ apply edge contraction on $e$
                \EndWhile
                \State $(G_1,c_1) \gets \Call{Merge\_Nodes}{G'}$
                \State $(G_2,c_2) \gets \Call{Merge\_Nodes}{G'}$
                \State \Return $\min_{c_i}\{(G_1,c_1),(G_2,c_2)\}$
		\EndFunction
	\end{algorithmic}
\end{algorithm}

\subsection{Circuit Cutting Algorithm}\label{sec:algorithm}
Based on the heuristic cost function, we are able to assess the routing overheads of a cutting solution.
In another word, we enable hardware-aware circuit cutting.
With such information, a cutting algorithm that takes both number of cuts and routing overhead into account can now be properly designed.

\noindent\textbf{Overview.} Our circuit cutting algorithm is given in Algorithm~\ref{rand}. 
It utilizes random edge contraction to merge nodes (qubits) into subgraphs (subcircuits)~\cite{kargerandstein},  
recursively samples a variety of cutting solutions,
and returns the solution with minimal cost $c$. 
Several constrains are applied to fit the algorithm into the circuit cutting scenario.
For size limitation on resulting subgraphs (maximum number of qubits in a subcircuit)~\cite{baker2020}, our algorithm sets constrain to avoid contraction that leads to merging into an oversize subgraph. 
% The cost $c$ is a combination of cuts number and graph edit distance, the coefficient $\alpha$ controls the trade-off between these two. % objectives. 

\noindent\textbf{Algorithm Details.} The major part of algorithm is \Call{Merge\_Nodes}{} function. Starting from the original qubit interaction graph with $n$ nodes, it randomly merges nodes into subgraphs. 
It keeps merging until the graph size reaches $\sqrt{n}$, then the function uses the current graph as input and invokes two recursive calls, each have a different merging to cover different solutions. 
The recursion keeps going until terminate condition ($E_{cst} = \varnothing$) is reached, then we get the solution subgraphs representing subcircuits. 
When returning from the function, it chooses the recursive calls that returns the solution with lower cost.

\subsection{Complexity Analysis.} 
According to the theory in \textit{Karger-Stein Algorithm}~\cite{kargerandstein}, the full program can be repeated for $O(log^2(n_q))$ times to achieve an error probability within $O(1/n_q)$ ($n_q$ is number of qubits). Running the program once need $O(n_q^2log(n_q))$ time, so the overall complexity is $O(n_q^2log^3(n_q))$. In comparison with the MIP solver used in CutQC, which have a $O(n_g!)$($n_g$ is number of 2-qubit gates, and $!$ is the factorial), our circuit cutting approach can be more scalable. 

\section{Classical Postprocessing}
\begin{figure}[ht]
    \centering
    \begin{subfigure}[b]{0.84\columnwidth}
        \centering
        \includegraphics[width=1\columnwidth]{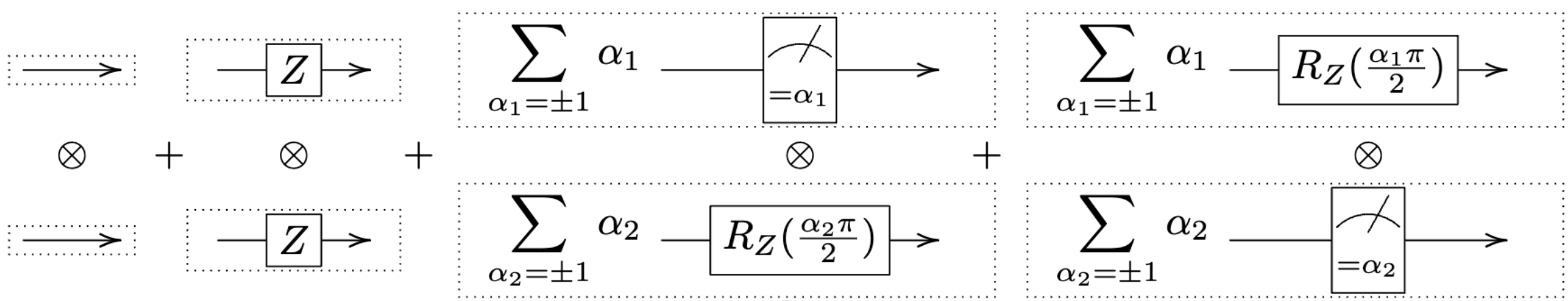}
    \end{subfigure}
    % \hfill
    \begin{subfigure}[b]{0.13\columnwidth}
        \centering
        \includegraphics[width=1\columnwidth]{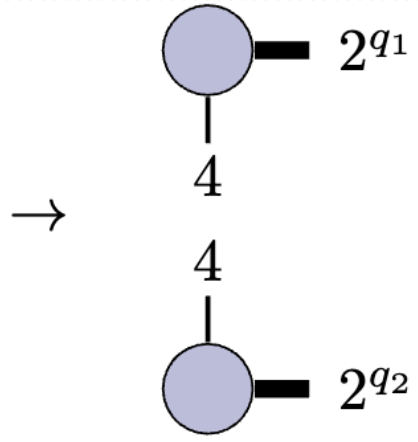}
    \end{subfigure}
    \caption{Classical postprocessing: merging two subcircuits result using tensor contraction. The notation on the right is a tensor network notation, representing the contraction of these two subcircuit result.}
    \label{fig:kronecker}
\end{figure}

\paragraph{\textbf{Optimizing Recombination Protocol.}} Classical postprocessing %is responsible for reconstructing 
reconstructs the result of the original circuit by merging  subcircuits result using Kronecker products.
Specifically, the basic protocol to merge results of two subcircuits is defined in Fig.~\ref{fig:v2qg}.
Note that the third term in the protocol will expand into $2\times4$ sub-terms (each corresponds to a Kronecker product) in respect to each value combination of the coefficients $\alpha_1 \alpha_2$.
Due to our observation that each coefficient $\alpha_i$ is only related to one side of the result in these sub-terms, we can split the combined coefficient $\alpha_1\alpha_2$ on each side, and the following Kronecker product would do recombination for them.
Thus, on each side of the result, the coefficient will calculate with subcircuit results ahead of Kronecker product, shrinking 4 sub-terms into one. Our preprocessing on the coefficients improve the number of Kronecker products of each merging from 10 ($=2+2\times4$) to 4, shown in Fig.~\ref{fig:kronecker}.

\paragraph{\textbf{Acceleration with Tensor Contraction.}} \todo{Furthermore, the merging operation in Fig.~\ref{fig:kronecker} is identical to the operation of tensor contraction. 
Kronecker product is a special version of tensor product when we group the $q$ indices each with the 2-dimension, into one single index with $2^q$-dimension (q=num of qubits). Assume that in both upper and lower side of Fig.~\ref{fig:kronecker}, the 4 result vectors (each with $2^{q_1}$ or $2^{q_2}$-dimension) can be seen as tensor in $4\times2^q$ shape, the whole merging operation is abstracted into the tensor contraction on the right of Fig.~\ref{fig:kronecker}. Correspondingly, if there are $n$ gate cuts between two subcircuits, the result tensor of each side would be in shape $4_1\times\dots\times4_n\times2^q$, who has $n$ legs with 4-dimension and one leg with $2^q$-dimension in tensor network notation. 
With such abstraction, we can translate the classical postprocessing problem into a tensor network %language for 
to searching a better solution.
%After transforming the classical postprocessing process into a tensor network, 
After translation,
% we are able to leverage the specific tools for find optimal contraction path, as well as the order of Kronecker products. 
% The optimal contraction path will form a tree structure, with an example given in Fig.~\ref{fig:contrationtree}. From the comparison between \textit{greedy-subcircuits-order} and our approach, it is obvious that our approach benefits from optimal contraction tree by reducing the Kronecker products and increasing the parallelism. 
we use NVIDIA cuTensorNet framework \cite{cuquantum} to enable an efficient classical postprocessing.}
% The cuTensorNet provides great acceleration and parallelism on GPUs, estimating a running time within 20 minutes for original circuit size up to 100 qubits.

\section{Evaluation}
Herein we are trying to answer if the proposed hardware-aware circuit cutting framework improves the state-of-the-art in terms of number of cuts, circuit depth, and fidelity, on a set of major quantum benchmarks, on real quantum hardware.
\todo{Furthermore, in Section~5.3, we test the scalability of our algorithm to determine its effectiveness with circuits of significantly larger scales. In Section 5.4, we enhance the QUEKO benchmark~\cite{queko} to generate circuits that have known optimal cutting solutions and qubit mappings, which we then use to study the optimality of our framework. Finally, in Section 5.5, we provide details on the classic postprocessing overhead when executed on a GPU.}
% We already disclose that our postprocessing stage is comparable to state-of-the-art. However, we are using GPUs to accelerate the compute process.

\begin{figure*}
    % \centering
    \begin{subfigure}[b]{0.75\columnwidth}
        \centering
        \includegraphics[width=1\columnwidth]{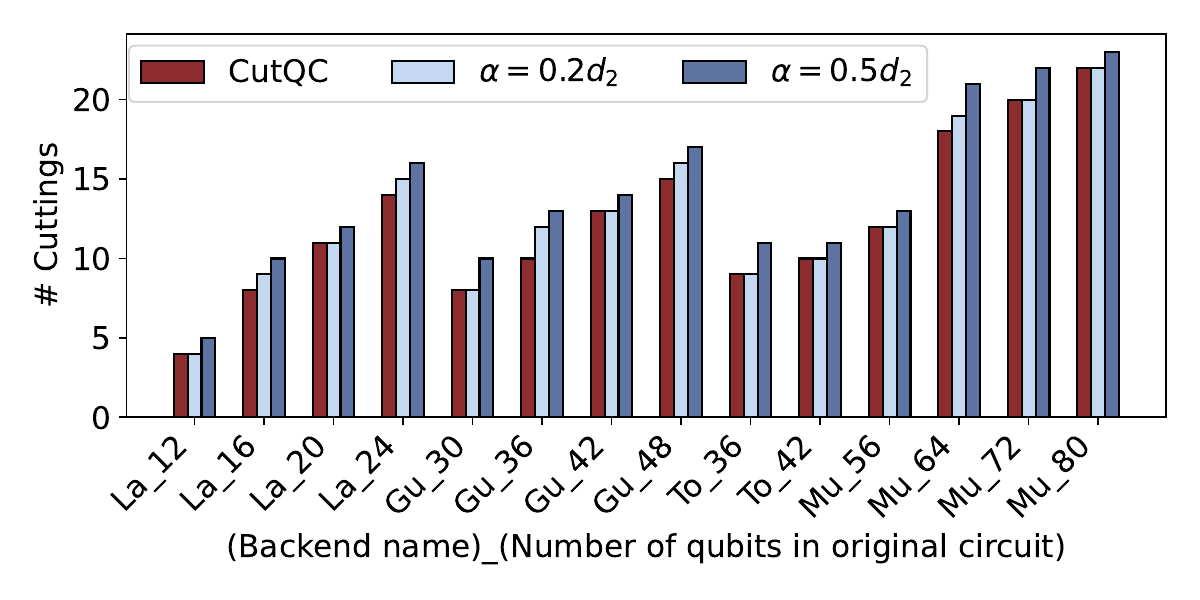}
        \caption{Cuts count for Supremacy circuit}
        \label{fig:s_cut}
    \end{subfigure}
    % \hfill
    \begin{subfigure}[b]{0.75\columnwidth}
        \centering
        \includegraphics[width=1\columnwidth]{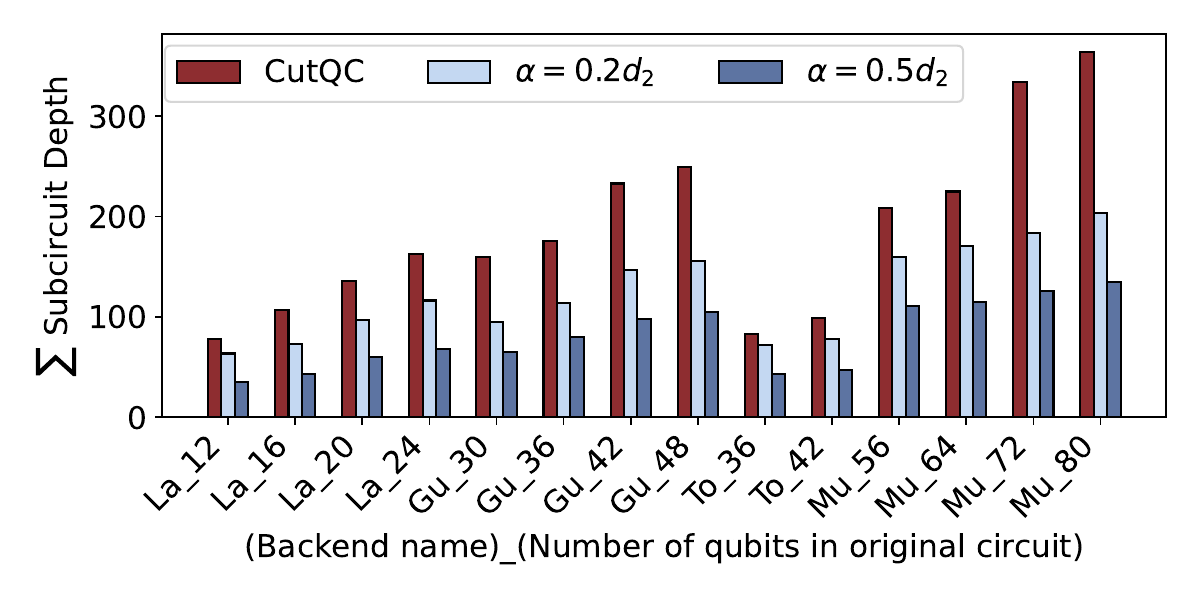}
        \caption{Depth count for Supremacy circuit}
        \label{fig:s_depth}
    \end{subfigure}
    \begin{subfigure}[b]{0.42\columnwidth}
        \centering
        \includesvg[width=1\columnwidth]{eval/fidelity_supremacy.svg}
        \caption{Relative fidelity of\\ Supremacy circuit}
        \label{fig:s_fidelity}
    \end{subfigure}

    \begin{subfigure}[b]{0.75\columnwidth}
        \centering
        \includesvg[width=1\columnwidth]{eval/evaluation_aqft_cut.svg}
        \caption{Cuts count for AQFT circuit}
        \label{fig:a_cut}
    \end{subfigure}
    \begin{subfigure}[b]{0.75\columnwidth}
        \centering
        \includesvg[width=1\columnwidth]{eval/evaluation_aqft_depth.svg}
        \caption{Depth count for AQFT circuit}
        \label{fig:a_depth}
    \end{subfigure}
    \begin{subfigure}[b]{0.42\columnwidth}
        \centering
        \includesvg[width=1\columnwidth]{eval/fidelity_aqft.svg}
        \caption{Relative fidelity of AQFT circuit}
        \label{fig:a_fidelity}
    \end{subfigure}

    \begin{subfigure}[b]{0.75\columnwidth}
        \centering
        \includesvg[width=1\columnwidth]{eval/evaluation_hwea_cut.svg}
        \caption{Cuts count for QAOA circuit}
        \label{fig:h_cut}
    \end{subfigure}
    \begin{subfigure}[b]{0.75\columnwidth}
        \centering
        \includesvg[width=1\columnwidth]{eval/evaluation_hwea_depth.svg}
        \caption{Depth count for QAOA circuit}
        \label{fig:h_depth}
    \end{subfigure}
    \begin{subfigure}[b]{0.42\columnwidth}
        \centering
        \includesvg[width=1\columnwidth]{eval/fidelity_hwea.svg}
        \caption{Relative fidelity of QAOA circuit}
        \label{fig:h_fidelity}
    \end{subfigure}

    \begin{subfigure}[b]{0.75\columnwidth}
        \centering
        \includesvg[width=1\columnwidth]{eval/evaluation_local_cut.svg}
        \caption{Cuts count for Ising circuit}
        \label{fig:l_cut}
    \end{subfigure}
    \begin{subfigure}[b]{0.75\columnwidth}
        \centering
        \includesvg[width=1\columnwidth]{eval/evaluation_local_depth.svg}
        \caption{Depth count for Ising circuit}
        \label{fig:l_depth}
    \end{subfigure}
    \begin{subfigure}[b]{0.42\columnwidth}
        \centering
        \includesvg[width=1\columnwidth]{eval/fidelity_local.svg}
        \caption{Relative fidelity of Ising circuit}
        \label{fig:l_fidelity}
    \end{subfigure}
    
    \caption{Evaluation. Row 1,2,3,4 are result for Supremacy, AQFT, QAOA, Ising benchmark respectively. Column 1 is the number of cuts, column 2 is sum of depth of subcircuits, column 3 is the relative fidelity. In the X-axis, La\_n, Gu\_n, To\_n, Mu\_n stand for IBM chip Lagos, Guadalupe, Tokyo, Mumbai running n-qubit size original circuit. It is notably that three groups of data of CutQC is blank for the AQFT benchmark, because solution is not feasible. }
    \label{fig:eval}
\end{figure*}

\subsection{Evaluation Methodology}\label{sec:evaluation_methodology}

\paragraph{\textbf{Framework Implementation:}} \todo{
%ORIGINAL We use the NetworkX~\cite{networkx} package to implement our circuit cutting algorithm, and use Qiskit~\cite{Qiskit} QuantumCircuit to partition and generate the subcircuits. 
We extended Qiskit~\cite{Qiskit}, we adopted the NetworkX~\cite{networkx} Python package to implement our circuit cutting algorithm, and use Qiskit's QuantumCircuit to partition and generate the subcircuits.
To run subcircuits on real-hardware, we use Qiskit transpiler for qubit mapping and routing, and selected the SABRE layout method~\cite{sabre} because of its speed and scalability.}

\paragraph{\textbf{Algorithm Configuration: }} \todo{We do provide an approach to decide the value of parameter $\alpha$ automatically, integrating the \textit{gate density} feature~\cite{queko}. Suppose the original circuit has $N$ logical qubits, $M_2$ two-qubit gates, and $l$ as the length of longest dependency chain, then the two-qubit gate density is $d_2 = 2 M_2 / (N \cdot l)$. We observe from experiments that multiples of $d_2$ are a good fit for the value of $\alpha$. Thus, we use two configurations in our evaluation: $\alpha = 0.2d_2$ and $\alpha = 0.5d_2$.}

% the machine I borrowed from Tencent
\paragraph{\textbf{Experiment Platform:}} \todo{%Our experiments are performed on a CentOS 7 server featuring an Intel Xeon Platinum 8255C CPU (320 GB memory) with 80 CPU cores at 2.5 GHz, and a NVIDIA V100 GPU with 5120 CUDA cores at 1.2 GHz. The Python version is 3.10.12.
Our experiments are performed on a CentOS 7 server featuring dual Intel Xeon Platinum 8255C CPUs for a total of 48 cores/96 threads at 2.5 GHz, 384 GB of RAM, and a NVIDIA V100 GPU with 5120 CUDA cores at 1.2 GHz. The Python version is 3.10.12.}

\paragraph{\textbf{Hardware Backends:}} We use four IBM quantum computers (free access with Open Plan~\cite{IBMpricing}): 7-qubit IBM Lagos, 16-qubit IBM Guadalupe, 27-qubit IBM Mumbai, and 127-qubit IBM Brisbane as backends. These hardware are all heavy-hexagon structure, so we add a fictitious 20-qubit IBM Tokyo (retired) backend, which has square structure, to prove the generalizability of our algorithm for different hardware layouts.
The original circuits and subcircuits are run from 32K to 1M shots, depending on the circuit size and sampling overhead.

\paragraph{\textbf{Benchmarks:}} We benchmark and compare with state-of-the-art our approach on four quantum algorithms that show potential for quantum advantages according to previous research:
\begin{enumerate}[]
\item \textit{Supremacy:} Google random circuit for quantum supremacy\cite{benchmark_supremacy};
\item \textit{AQFT:} Approximate Quantum Fourier Transform\cite{benchmark_aqft};
\item \textit{QAOA:} Hardware efficient ansatz for Quantum Approximate Optimization Algorithm\cite{benchmark_qaoa};
\item \textit{Ising}: 2-local Hamiltonian Simulation with Ising Model \cite{2qan}.
\end{enumerate}%
%\paragraph{\textbf{Size of original circuit:}}
We set the size of original circuit from $1.5\times$ to $4\times$ of the physical qubits available on the selected hardware backends.

\paragraph{\textbf{Metrics:}} We use the following metrics to evaluate the overheads introduced in Sect.~\ref{sec:overheads}.
\begin{enumerate}
    \item \textit{Number of cuts.} To establish a fair comparison that mitigate the impact of using different classical computing platforms and quantum hardware, we directly choose the number of cuts $n$ as metric, to evaluate the sampling overhead $O(\gamma^{2n})$ and postprocessing overhead $O(4^n)$.

    \item \textit{Sum of subcircuits' depth.} The routing overhead of a subcircuit is reflected on its depth. Because the same original circuit will have different cutting solution in different frameworks, each of the subcircuit's depth is unmatched and incomparable. However, we can compare the sum of the depths of all subcircuits.

    \item \textit{Relative fidelity on quantum computer.} We select the IBM Lagos (7-qubit) and IBM Guadalupe (16-qubit) quantum computer as the hardware backend to run circuit knitting, and compare its relative fidelity with directly running the original circuit on IBM Mumbai (27-qubit).
\end{enumerate}

% In the Experiment 1 that compares number of cuts and subcircuits depth, all the hardware layout are included. For the Experiment 2, which is comparison on real device, we focus on the three heavy-hexagon layouts of IBM's quantum computers. 

\subsection{Results}
Our evaluation results are shown in Fig.~\ref{fig:eval}.
We use the state-of-the-art CutQC as a baseline, instead of its improved version ScaleQC~\cite{scaleqc}, because the source code of ScaleQC is still not available. Specifically, we test our framework with two configurations: $\alpha = 0.2d_2$ and $\alpha = 0.5d_2$ as introduced in Section~5.1.
% \begin{enumerate}
%     \item \textit{CutQC}: is a wire cutting approach with optimal solution of minimal number of cuttings on qubit wires.
%     \item \textit{Gcut\_optimal}: is a gate cutting approach with near-optimal solution of minimal number of cuttings on qubit gates, meanwhile trying the best to minimize number of SWAPs.
%     \item \textit{Gcut\_HW-aware}: is a gate cutting approach with trading off bet
% \end{enumerate}
% In the \textit{trade\_off} configuration, we set value of $\alpha$ as 0.25, 0.2, 0.5 and 0.3 respectively for Supremacy, AQFT, QAOA and Ising. 
With IBM quantum computers as the backends, three metrics are evaluated in four benchmarks. The x-axis in Fig.~\ref{fig:eval} is in the format of [Backend Name]\_[Original circuit qubits].

\todo{For Metrics (1) and (2), the scale of the original circuits is limited to 80 qubits, constrained by the ability of baseline framework~\cite{cutqc}. For Metric (3), the calculation of relative fidelity depends on noise-free results obtained from simulations; therefore, the limitation is limited by the capability of current quantum simulators, which is approximately 33 qubits.}

It is notably that in AQFT benchmark, CutQC have "solution not feasible" during the run in 16\_40, 16\_48, and 27\_64, so the corresponding results are left blank in Figure~\ref{fig:a_cut} and ~\ref{fig:a_depth}.

\paragraph{\textbf{Cuts and Depth.}}
Compared to CutQC as baseline, our framework with the $\alpha = 0.2d_2$ configuration increases the number of cuts by 3\% on average, but reduces the subcircuits depth from 25\% to 64\% (34\% on average; 43\%, 47\%, 29\%, 17\% respectively for benchmark Supremacy, AQFT, QAOA, and Ising). The $\alpha = 0.5d_2$ configuration increases the number of cuts by 13\% on average compared to baseline, but it reduces the subcircuits depth from 25\% to 64\% (48\% on average; 57\%, 53\%, 48\%, 36\% respectively for benchmark Supremacy, AQFT, QAOA, and Ising). \todo{Based on these results, our framework effectively reduces circuit depth while accommodating a manageable increase in the number of cuts, demonstrating scalability to larger circuit sizes.}

\paragraph{\textbf{Relative Fidelity.}}
For relative fidelity on quantum computers, the $\alpha = 0.2d_2$ and $\alpha = 0.5d_2$ configurations have an average result of $1.54\times$ and $1.73\times$ for Supremacy, $1.13\times$ and $1.1\times$ for AQFT, $1.55\times$ and $1.98\times$ for QAOA, $1.5\times$ and $1.67\times$ for Ising, and it's $1.39\times$ and $1.65\times$ totally for all benchmarks. Compared to CutQC, we have an improvement up to 231\%. Our framework have relative fidelity > 1 on almost every benchmark, while CutQC have negative results in part of the benchmarks. Therefore, we identified that our framework enhance the practicality of circuit knitting by reducing routing overhead.

\begin{figure}[b]
    \begin{subfigure}[b]{0.35\columnwidth}
        \centering
        \includesvg[width=1\columnwidth]{eval_scale/eval_supremacy_cut.svg}
    \end{subfigure}
    \begin{subfigure}[b]{0.35\columnwidth}
        \centering
        \includesvg[width=1\columnwidth]{eval_scale/eval_supremacy_depth.svg}
    \end{subfigure}
    
    \begin{subfigure}[b]{0.35\columnwidth}
        \centering
        \includesvg[width=1\columnwidth]{eval_scale/eval_aqft_cut.svg}
    \end{subfigure}
    \begin{subfigure}[b]{0.35\columnwidth}
        \centering
        \includesvg[width=1\columnwidth]{eval_scale/eval_aqft_depth.svg}
    \end{subfigure}

    \begin{subfigure}[b]{0.35\columnwidth}
        \centering
        \includesvg[width=1\columnwidth]{eval_scale/eval_hwea_cut.svg}
    \end{subfigure}
    \begin{subfigure}[b]{0.35\columnwidth}
        \centering
        \includesvg[width=1\columnwidth]{eval_scale/eval_hwea_depth.svg}
    \end{subfigure}

    \begin{subfigure}[b]{0.35\columnwidth}
        \centering
        \includesvg[width=1\columnwidth]{eval_scale/eval_local_cut.svg}
    \end{subfigure}
    \begin{subfigure}[b]{0.35\columnwidth}
        \centering
        \includesvg[width=1\columnwidth]{eval_scale/eval_local_depth.svg}
    \end{subfigure}
    
    \caption{Evaluation. Row 1,2,3,4 are result for Supremacy, AQFT, QAOA, Ising benchmark respectively. Column 1 is the number of cuts, column 2 is sum of depth of subcircuits. In the X-axis, Br\_n stand for IBM chip Brisbane (127-qubit) running n-qubit size original circuit.}
    \label{fig:eval_scale}
\end{figure}

\subsection{Scaling Up in Circuit Size}
\todo{In Section~5.2 the evaluation of \#cuts and subcircuits depth are limited to circuits up to 80 qubits, constrained by the capabilities of baseline framework~\cite{cutqc}. In complementary to this, experiment results on larger circuits (ranging from 100 to 1000 qubits) are presented in Figure~\ref{fig:eval_scale}, using IBM Brisbane (127-qubit) for the hardware topology. Our observations indicate that while setting $\alpha=0.5k$ results in a slight increase in \#Cuts compared to $\alpha=0.2k$, it considerably reduces the depth of subcircuits. These results effectively demonstrate that as circuit size increases to larger scale (1000 qubits), our algorithm can still successfully find suitable trade-off solutions aligned with the Pareto frontier.}

\subsection{Optimality Study}
\begin{figure}[h]
    \centering
    \includegraphics[width=0.8\columnwidth]{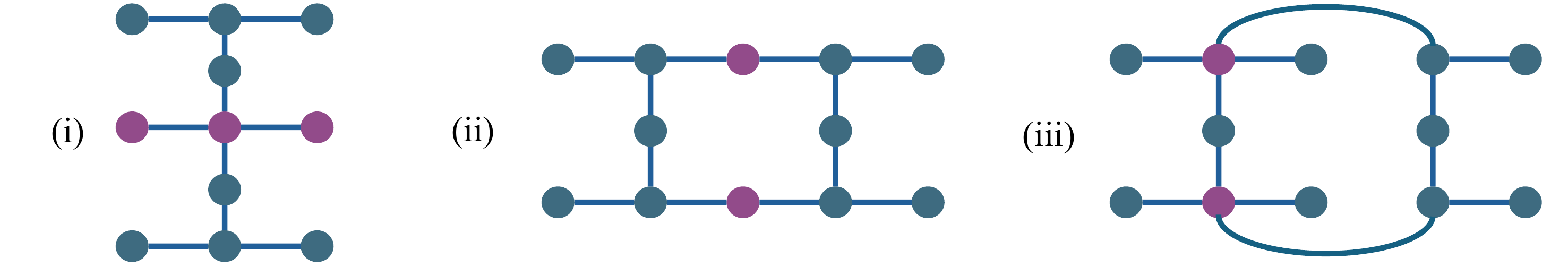}
    \caption{The topology of virtual devices created by connecting multiple IBM Lagos chips.}
    \label{fig:queko}
    \vspace{-1mm}
\end{figure}

\todo{We have enhanced the QUEKO \cite{queko} to evaluate the optimality of our algorithm. Specifically, we created several "virtual quantum hardware" by connecting together multiple instances of IBM Lagos topology, illustrated in Fig.~\ref{fig:queko}. Based on these virtual device topology, we employed QUEKO to generate logical circuits with known optimal cutting and mapping solutions. Subsequently, we tested our framework ($\alpha=0.2k$) on these circuits and compared the results with the baseline. The data presented in Table\ref{table:optimal} show that our approach is more likely to find the optimal solution when considering multiple overheads.}

\begin{table}[h]
  \small
  \centering
  \caption{Optimality Study. The result data is presented as the form of [CutQC | Our Work].}
  \begin{tabular}{|c|c|c|}
    \hline
    Original Circuit & \#Cuts &  Subcircuits Depth Added \\
    Generated From   &        &  Compared to Optimal Solution \\
    \hline
    (i)   & 1 | 1 & 0\% | 0\% \\
    \hline
    (ii)  & 2 | 2 & 13\% | 0\% \\
    \hline
    (iii) & 2 | 2 & 21\% | 0\% \\
    \hline
  \end{tabular}
  \vspace{-1mm}
  \label{table:optimal}
\end{table}

\subsection{Classical Postprocessing}
\begin{figure}[h]
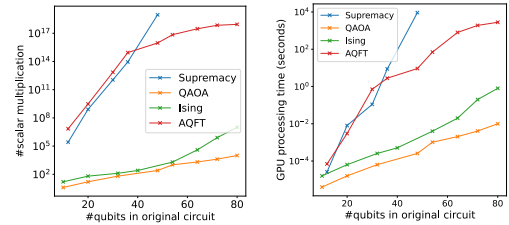

    % \centering
    \begin{subfigure}[b]{0.4\columnwidth}
        \centering
        \includesvg[width=1\columnwidth]{postprocessing_cost.svg}
    \end{subfigure}
    % \hfill
    \begin{subfigure}[b]{0.4\columnwidth}
        \centering
        \includesvg[width=1\columnwidth]{postprocessing_GPU.svg}
    \end{subfigure}
    \caption{Overheads associated with result recombination in postprocessing. (a) illustrates the number of scalar multiplications to be performed in postprocessing, while (b) shows the postprocessing time on a NVIDIA V100 GPU.}
    \label{fig:postprocessing}
\end{figure}

\todo{%ORIGINAL As the classical postprocessing approach described in Section~4, we test its performance on our benchmarks. For QAOA and Ising benchmarks, we only need to recombine the expectation value of each subcircuits, instead of recombining the full measurement results, so the overheads are quite low, compared with other algorithms. For Approximate-QFT benchmark, we adopt the "Dynamic Definition" approach purposed in \cite{cutqc} when $qubits \geq 50$, to reduce the overhead. The results are comparable with \cite{cutqc}.
We tested the performance of our classical postprocessing approach, described in Section~4, on the benchmarks we used for the evaluation (see Section~5.1). The results are shown in Fig.~\ref{fig:postprocessing}.
For QAOA and Ising benchmarks, we only recombine the expectation value of each subcircuits, instead of recombining the full measurement results, so the overheads are quite low compared with other algorithms.
For Approximate-QFT benchmark, we adopt the "Dynamic Definition" approach purposed in \cite{cutqc} when $qubits \geq 50$, to reduce the overhead, while we don't apply any optimization for the Supremacy benchmark. %, which show the higher postprocessing time.
Results follow the same trends as previous works, our postprocessing approach seem faster than~\cite{cutqc} -- likely, because of using a GPU vs CPU.
}

\section{Related Works}
CutQC\cite{cutqc} is a framework that cuts a large quantum circuit in subcircuits and execute them on (small) quantum computers using the \textit{wire cut} protocol~\cite{peng2020}. 
ScaleQC~\cite{scaleqc} is an extension of CutQC's postprocessing module, integrating tensor network contraction specific for \textit{wire cut}. Our work focuses on \textit{gate cut} protocol.
Qiskit Circuit Knitting Toolbox (CKT)~\cite{ckd} provides out-of-the-box circuit knitting protocols, but the core cutting solution have to be instructed manually by the user. Moreover, CKT's postprocessing only calculates the result of observable expectation, thus only supports applications like VQE and QAOA.
Baker et al. propose a time-sliced circuit cutting\cite{baker2020}, while Tomesh et al. \cite{tomesh2021divide} design a \textit{wire cut} framework for QAOA programs.
No previous work integrates hardware layout information during circuit cutting, like we propose. %d by this work.

FCM \cite{mo2024fcm} utilizes wire cutting to improve the fidelity of Measurement Based Quantum Computing (MBQC), while our framework is compatible with various type of quantum backends.
TopGen \cite{cheng2022topgen} generates topology-aware subcircuits for VQA algorithm. 
JigSaw \cite{jigsaw} and Quixote~\cite{jang2023quixote} are error mitigation frameworks that measure only part of the qubits and reconstructs the result with higher fidelity; similarly, Li et al. \cite{qutracer_jiliu1,jiliu2} leverage circuit cutting to enhance the effectiveness of error mitigation techniques, while we focus on optimizing circuit cutting and compilation themselves to improve the result.
Bandic et al. utilize graph theory-based metrics to characterize quantum circuits\cite{characterization}.

\section{Conclusion}

Developing methods to fully exploit current NISQ devices with their limited number of qubits has been an active area of research.
In this work, we propose a hardware-aware circuit knitting framework with graph similarity optimization, to facilitate the identification of better cutting schemes that can significantly reduce circuit depth.
We fully implemented the entire framework and performed evaluations using real quantum processors.
Our results surpass those of approaches that perform cutting and compilation independently, pioneering pathways for practical circuit knitting. 
As quantum hardware continues to advance, our techniques will be invaluable for future distributed quantum computing scenarios. 

\paragraph{\textbf{Acknowledgments.}}
Xiangyu Ren thanks Dr. Yi-Cong Zheng for his precious early guidance on the theory of circuit knitting. We thank Dr. Shengyu Zhang and Tencent Quantum Lab for the generous suggestion on paper writing and revising.

\bibliographystyle{plain}
\bibliography{references}

\end{document}